\documentclass{article}

\PassOptionsToPackage{numbers, compress}{natbib}

\usepackage[final]{neurips_2024}

\usepackage[utf8]{inputenc} 
\usepackage[T1]{fontenc}    
\usepackage{hyperref}       
\usepackage{url}            
\usepackage{booktabs}       
\usepackage{amsfonts}       
\usepackage{nicefrac}       
\usepackage{microtype}      
\usepackage{xcolor}         

\usepackage{adjustbox}
\usepackage{diagbox}
\usepackage{multirow}
\usepackage{makecell}
\usepackage{subfigure}

\usepackage{times}
\usepackage{graphicx}
\usepackage{epstopdf} 
\usepackage{epsfig}
\usepackage{amsmath}
\usepackage{amssymb}
\usepackage{colortbl}
\usepackage{caption}
\usepackage{ragged2e}
\usepackage{float}
\usepackage{booktabs}
\usepackage{bbding}

\title{IndusGCC: A Data Benchmark and Evaluation Framework for GUI-Based General Computer Control in Industrial Automation}
\author{Xiaoran Yang\textsuperscript{\rm 1,2,\thanks{Work conducted during visiting in IE Dept., CUHK, $^\dagger$ Project Lead, $^\ddagger$ Corresponding Author} }, 
Yuyang Du\textsuperscript{\rm 2,$\dagger$},
Kexin Chen\textsuperscript{\rm 3,$\dagger$}, 
Soung Chang Liew\textsuperscript{\rm 2,$\ddagger$},
Jiamin Lu\textsuperscript{\rm 4}, \\
\textbf{Ziyu Guo\textsuperscript{\rm 3}},
\textbf{Xiaoyan Liu\textsuperscript{\rm 2}},
\textbf{Qun Yang\textsuperscript{\rm 2}},
\textbf{Shiqi Xu\textsuperscript{\rm 2}},
\textbf{Xingyu Fan\textsuperscript{\rm 3}},
\textbf{Yuchen Pan\textsuperscript{\rm 2}},
\textbf{Taoyong Cui\textsuperscript{\rm 3}},\\
\textbf{Hongyu Deng\textsuperscript{\rm 2}},
\textbf{Boris Dudder\textsuperscript{\rm 1}},
\textbf{Jianzhang Pan\textsuperscript{\rm 5,6}},
\textbf{Qun Fang\textsuperscript{\rm 5,6}},
\textbf{Pheng Ann Heng\textsuperscript{\rm 3,$\ddagger$}}
\vspace{0.125cm}\\
\textsuperscript{\rm 1}Department of Computer Science, the University of Copenhagen\\
\textsuperscript{\rm 2}Department of Information Engineering, the Chinese University of Hong Kong\\
\textsuperscript{\rm 3}Department of Computer Science Engineering, the Chinese University of Hong Kong\\
\textsuperscript{\rm 4}The First Affiliated Hospital of Xiamen University, School of Medicine, Xiamen University\\
\textsuperscript{\rm 5}Institute of Intelligent Chemical Manufacturing and iChemFoundry Platform,\\ZJU-Hangzhou Global Scientific and Technological Innovation Center\\
\textsuperscript{\rm 6}Institute of Microanalytical Systems, Department of Chemistry, Zhejiang University
\vspace{0.125cm}\\
\textit{Corresponding Emails: soung@ie.cuhk.edu.hk, pheng@cse.cuhk.edu.hk}
}

\begin{document}

\maketitle

\vspace{-0.3cm}
\begin{abstract}
\vspace{-0.1cm}
As Industry 4.0 progresses, flexible manufacturing has become a cornerstone of modern industrial systems, with equipment automation playing a pivotal role. However, existing control software for industrial equipment, typically reliant on graphical user interfaces (GUIs) that require human interactions such as mouse clicks or screen touches, poses significant barriers to the adoption of code-based equipment automation. Recently, Large Language Model-based General Computer Control (LLM-GCC) has emerged as a promising approach to automate GUI-based operations. However, industrial settings pose unique challenges, including visually diverse, domain-specific interfaces and mission-critical tasks demanding high precision. This paper introduces IndusGCC, the first dataset and benchmark tailored to LLM-GCC in industrial environments, encompassing 448 real-world tasks across seven domains, from robotic arm control to production line configuration. IndusGCC features multimodal human interaction data with the equipment software, providing robust supervision for GUI-level code generation. Additionally, we propose a novel evaluation framework with functional and structural metrics to assess LLM-generated control scripts. Experimental results on mainstream LLMs demonstrate both the potential of LLM-GCC and the challenges it faces, establishing a strong foundation for future research toward fully automated factories. Our data and code are publicly available at: \href{https://github.com/Golden-Arc/IndustrialLLM}{https://github.com/Golden-Arc/IndustrialLLM.}

\end{abstract}

\vspace{-0.25cm} \section{Introduction}\vspace{-0.25cm}
As the world enters the Industry 4.0 era, system automation has become indispensable for modern factories. In complex production environments, particularly in flexible manufacturing, machines often require dynamic reconfiguration to adapt to shifting production goals, supply constraints, or real-time operational feedback. In large-scale smart factories, engineers may perform tens of thousands of configuration tasks daily, such as setting robot motion parameters or adjusting networking device configurations. Significant human workload could be alleviated if these operations were automated and conducted in an unmanned manner, using control scripts generated by equipment management frameworks for system adjustments.

There have been some previous attempts in building specialized industrial software for automatic equipment control, such as typical Programmable Logic Controller (PLC) environments like Unity Pro \cite{unitypro}, IndraWorks \cite{indraworks}, and TwinCAT \cite{twincat}. These tools are designed to facilitate automation by leveraging scripting capabilities in manufacturing and plant operations. Yet, as flexible manufacturing continues to advance, these customized industrial software solutions have gradually revealed several shortcomings. Such software often suffers from lengthy development cycles, limited flexibility, and poor generalizability. Each new manufacturing technique typically demands additional software engineering efforts, including customized PLC configurations and the development of new control scripts. These tools, lacking the ability to scale across diverse industrial platforms, are rarely capable of autonomously adapting to new user interfaces or workflows without extensive re-engineering.

Recent investigations in Large Language Models (LLMs) have brought new possibilities in AI-assisted industrial device control. Leveraging the reasoning and generalization abilities of LLMs, prior works have developed LLM-driven coordinators for automating industrial device management \cite{LLMind} across diverse platforms, fine-tuned LLM agents for reliable control script generation \cite{LLMind2}, and efficient industrial information matching based on Retrieval-Augmented Generation (RAG) \cite{Chemist-X, ChemMiner}.

\begin{figure}[t]
\begin{center}
\captionsetup{font={footnotesize}}
\includegraphics[width=0.575\textwidth]{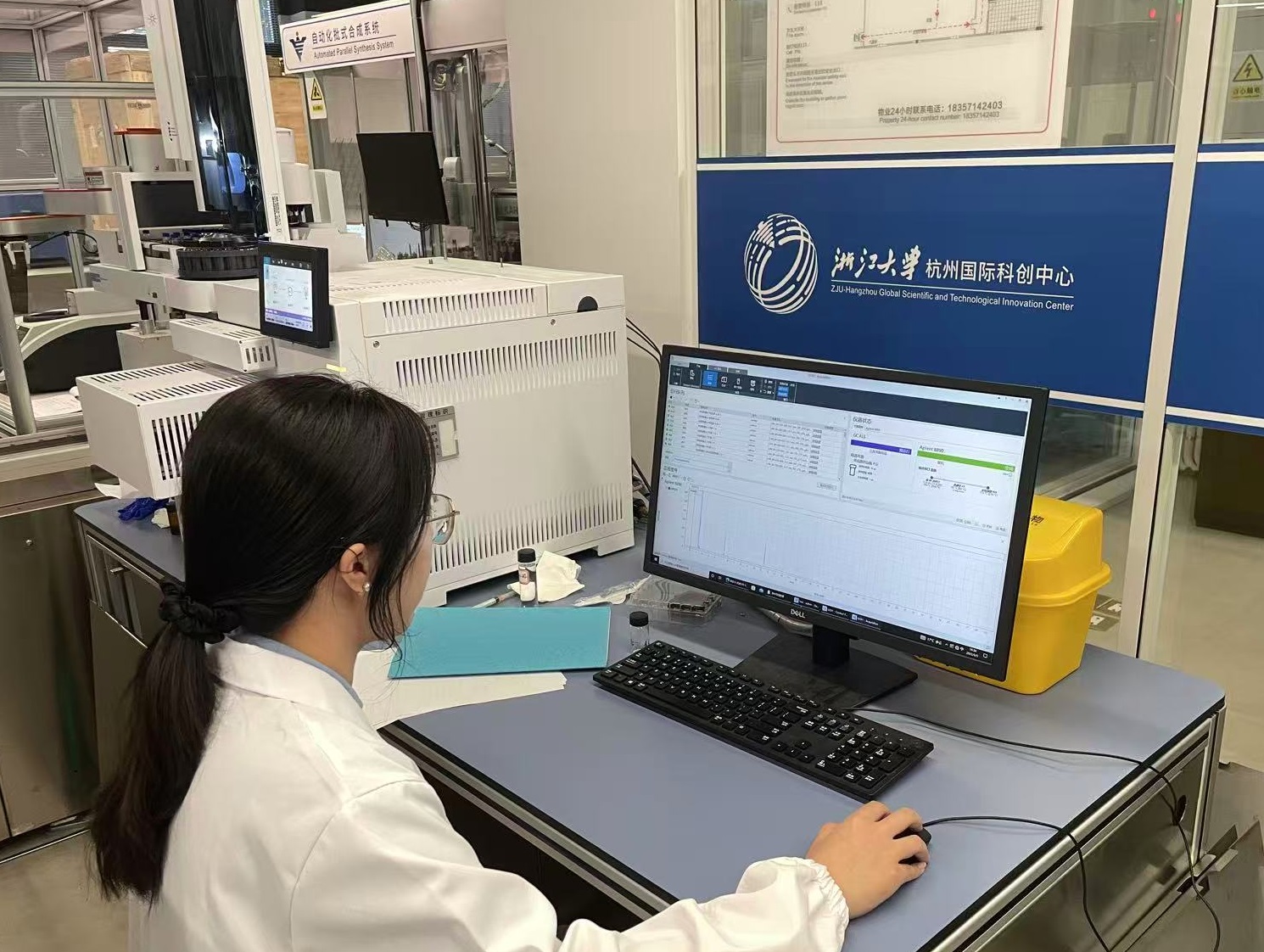}
\caption{A chemist controlling an automative synthesis system (on the left of the image, not fully recorded). This image is a typical example of the most frequently used equipment control method via computer I/O such as keyboards and mouse. \vspace{-0.6cm}}
\label{fig:intro}
\end{center}
\end{figure}

Despite the promising picture of LLM-assisted factory management, a common assumption in previous works \cite{LLMind, LLMind2, Chemist-X, ChemMiner} is the availability of programming-based control interfaces for industrial devices. However, in most factories today, it remains common for industrial computers to be operated via I/O devices, such as keyboards, mice, and touchscreens (see Fig. \ref{fig:intro} for example). These Graphical User Interface (GUI)-based configurations rely on skilled workers to manually interact with specialized software GUIs rather than machine-understandable programming interfaces, making LLM-supervised device automation particularly challenging.

The concept of LLM-empowered General Computer Control (GCC), which develops AI agents capable of interacting with arbitrary GUI-based software to perform tasks in a manner similar to human users, could fill the critical gap between the assumption in previous works \cite{LLMind, LLMind2, Chemist-X, ChemMiner} and the practical situation in real factories. However, prior research on LLM-GCC has primarily focused on navigation in everyday scenarios, such as interacting with webpage\cite{AutoWebGLM,WebCanvas,WEBLINX,WebShop,WebVoyager,MIND2WEB}, mobile device\cite{Mobile-Agent,Mobile-Agent-v2,SPA-Bench,AITW,androidlab,AndroidWorld}, or regular office software\cite{macOSWorld,windowsagentarena,worldgui,OSWorld,OmniACT}. These environments are typically structured, homogeneous, and rely on system-level representations like Document Object Model (DOM) trees. Industrial GUI environments, on the other hand, are far more complex. They are often visually diverse, closed-source, and highly domain-specific, featuring dynamic menus, real-time system feedback, and extensive parameter spaces. Additionally, operations in industrial settings frequently have mission-critical consequences, requiring precise execution and high reliability.

While LLM-GCC is both critical and challenging for industrial applications, existing research has made little progress in this domain. To date, there is not even a publicly available dataset specifically addressing LLM-GCC for industrial scenarios, let alone further exploration of LLM pre-training or post-training tailored to these applications.

Building on the above observations, this paper attempts to address the absence of an industrial GCC dataset. Through collaboration with real, operational factories, we collected a real-world, GUI-based dataset that captures workers' interactions while operating industrial equipment. This dataset, known as IndusGCC, is the first publicly available benchmark suite specifically designed for LLM-GCC in industrial settings, representing an important step towards fully automatic factory management.

Major contributions and key insights of this paper are summarized as follows.

\textbf{Introduction of a Large-Scale Dataset:} We present IndusGCC, a large-scale dataset encompassing 448 real-world tasks across seven industrial domains, including robotics control, mission-critical industrial networking, and manufacturing automation. Each task instance is paired with multimodal human interaction data that are synchronized in time, including screen recordings, mouse positions, mouse and keyboard events, and textual task descriptions.  This dataset is designed to provide rich supervision signals for GUI-level code generation and facilitate future research in this area.

\textbf{Comprehensive Evaluation Metrics:} We propose a set of evaluation metrics that closely align with real-world usability. Task success is measured based on functional code equivalence -- specifically, whether the LLM-generated control script produces the same system behavior as a gold-standard code reference, which represents how a human worker would operate the equipment. Additionally, we assess step completeness of the LLM-generated scripts using the Smith-Waterman algorithm to compare Abstract Syntax Tree (AST) sequences, capturing structural alignment between the LLM-generated code and the reference code.

\textbf{Benchmarking Mainstream LLMs:} Using the novel dataset and evaluation methods, we benchmark several mainstream LLMs to assess their performance in industrial control script generation. The results demonstrate the feasibility of LLM-GCC in industrial contexts while highlighting notable performance gaps. These gaps reveal the need for improvements in spatial perception, temporal understanding, and behavior planning in current models. We believe this benchmarking effort provides a meaningful foundation for future research in this area.

\vspace{-0.225cm} \section{Related work} \vspace{-0.225cm}
\textbf{Dataset and Benchmark for LLM-GCC:} 
A growing number of works have emerged in building benchmarks for LLM-GCC. Existing datasets vary significantly in terms of task type and observation space (i.e. data types that is allowed as the agent’s input).

Early attempts at constructing benchmarks for LLM-GCC largely focused on web navigation tasks, where well-structured environments like DOM Trees offered deterministic state representations \cite{WEBLINX,WebShop}. Over time, research expanded toward more diverse and realistic settings such as mobile device control, aiming to approximate the complexity of human-computer interaction in mobile operating systems \cite{AITW,androidlab,AndroidWorld}. In recent datasets, tasks go beyond simple navigation, increasingly encompassing daily-use general applications including document editing, video processing, gaming, and even software development \cite{SPA-Bench,macOSWorld,windowsagentarena,worldgui,AgentBoard,GAIA,Spider2-V,VideoWebArena}.

To support such complexity, the design of the GCC task’s observation space has also undergone significant evolution. Initial benchmarks provided only natural language prompts as input \cite{AgentBoard,GAIA,WorkArena++}, while later works introduced well-structured control representations such as DOM trees and Android Extensible Markup Language (XML) data to bridge the gap between high-level instructions and low-level UI elements \cite{WebShop,WebVLN}. Variants of the ``Set-of-Mark" (SOM) format \cite{SOM} became standard in many recent mobile- and web-based datasets due to their modularity and adaptability. However, these formats inherently rely on the availability of structured access, which limits their applicability to closed-source software environments. With the advent of multimodal LLMs (MLLMs), recent efforts have begun to fuse visual inputs -- screenshots or videos -- with SOM, as richer observation modalities \cite{WebCanvas,WebVoyager,MIND2WEB,androidlab,AndroidWorld,macOSWorld,windowsagentarena,worldgui,OSWorld,assistantbench,MMInA,VisualWebArena,WebArena}. Yet, most of these visual benchmarks continue to serialize perception through SOM-like abstractions rather than embracing fully end-to-end visual modeling.

IndusGCC pushes this frontier by addressing task setups where structured access is entirely absent, which is in fact a common case in closed-source industrial software that exposes neither DOM trees nor accessibility APIs for safety reasons. We adopt a minimal observation setup, where the agent must rely solely on raw screen pixels and natural language task instructions to perceive state and make decisions. This setting reflects a realistic and highly constrained industrial use case, where interface complexity and software opacity preclude structured interaction, and intelligent agents must operate purely from visual and linguistic grounding.

Equally important, high-quality industrial interaction data is exceptionally scarce: it must be captured through direct engagement with operational systems in real production environments. To address this fundamental gap, our dataset collects 448 authentic tasks, recorded across seven key application domains, each grounded in routine yet critical factory operations. These tasks include configuring wireless routers in network management rooms, designing signal flows for radio-frequency testbeds, performing robotic motion planning in robot calibration stations, optimizing manipulator trajectories in assembly-line teaching scenarios, inspecting protocol packets during industrial system diagnostics, programming weld sequences in automated welding workcells, and orchestrating reagent pipelines in chemical laboratories. This makes IndusGCC the first of its kind to support vision-based, instruction-driven control across such a broad and practically grounded set of real-world industrial contexts.

\textbf{Evaluation for LLM-GCC:} 
In recent studies, evaluation methods of LLM-GCC systems have progressed from early manual annotation schemes -- often used in small-scale studies or proof-of-concept datasets -- to more systematic and scalable automatic evaluation protocols \cite{Mobile-Agent,Mobile-Agent-v2,AITW}. Automatic evaluation is now standard across most recent benchmarks and can be broadly categorized into offline, interactive, and hybrid approaches. Offline evaluation compares model outputs to predefined ground truth without requiring execution in a live environment \cite{WEBLINX,MIND2WEB,OmniACT,GAIA}. Interactive evaluation executes model-generated actions within a real or simulated interface and determines success based on the resulting system state. Within interactive settings, some datasets operate in open environments (e.g., browsers or operating systems with uncontrolled variability), while others rely on customized mirrored environments designed to ensure reproducibility and safety \cite{AutoWebGLM,WebCanvas,WebShop,WebVoyager,SPA-Bench,AndroidWorld,macOSWorld,windowsagentarena,worldgui,OSWorld,Spider2-V,WorkArena++,VisualWebArena,WebArena}. Hybrid evaluation combines offline and interactive components -- often scoring intermediate reasoning steps or plans offline, while verifying final execution outcomes via interactive runs \cite{WebShop,androidlab,AgentBoard,VideoWebArena}.

In terms of evaluation metrics, task success rate has emerged as the most widely adopted and critical indicator. As a result-driven metric, it reflects the agent’s overall perception, planning, and decision-making competence across varying task types. To complement this binary outcome measure, several benchmarks introduce step-wise completion or efficiency scores -- quantifying the fraction of gold-standard steps reproduced by the agent -- which can reveal omissions or redundancies in generated control sequences \cite{WebCanvas,MIND2WEB,Mobile-Agent,Mobile-Agent-v2,SPA-Bench,AITW,AgentBoard,WebVLN}. Other metrics include action accuracy, such as the validity of mouse click coordinates \cite{androidlab}, and intermediate state correctness, which evaluates partial progress toward goals or correct identification of UI elements during execution \cite{WEBLINX,Mobile-Agent-v2,androidlab,GAIA,VideoWebArena,WebVLN,assistantbench,MMInA,VisualWebArena}.

Compared with prior works, our evaluation protocol introduces two key innovations: First, we employ richer and more fine-grained evaluation metrics beyond traditional task success rates. Each model prediction is represented as executable PyAutoGUI-based control code, enabling multiple dimensions of assessment. We adopt an LLM-as-judge mechanism to evaluate whether predicted and reference scripts are functionally equivalent -- based not merely on syntax but on their actual control effect. This also eliminates the need for researchers to manually configure or simulate execution environments. To further evaluate the internal structure of action sequences, we also compute step-level similarity and operation-level precision using a Smith-Waterman alignment algorithm over AST. This evaluation framework provides an interpretable and rigorous measure of outcome quality and procedural fidelity.

Second, IndusGCC addresses a challenge often overlooked in previous datasets: the problem of operational tolerance. In real industrial scenarios, some control actions -- especially those involving mouse positioning -- do not require pixel-perfect accuracy, but rather fall within tolerable spatial ranges. We annotate these tolerances explicitly IndusGCC, allowing the evaluation process to distinguish between semantically correct and functionally invalid actions. This enables models to better reason about control affordances and prevents unfair penalization of predictions that are practically acceptable but not identically matched. As a result, IndusGCC provides a realistic and robust foundation for measuring model performance under real-world constraints.

\vspace{-0.225cm} \section{Dataset and Baseline}\vspace{-0.225cm}
\subsection{Dataset Overview}\vspace{-0.15cm}
We construct the first large-scale industrial GUI operation dataset by collecting human interaction videos from seven real-world industrial scenarios, each of which reflects practical production needs and high operational complexity. These scenarios cover:

\begin{enumerate}
\item
Robotic arm control (precise motion parameter tuning and task execution by MOVEIT\cite{moveit}); 
\item
Network device configuration (e.g., access point setup, router management); 
\item
USRP simulation (software-defined radio configuration and signal processing)\cite{gnu,uhd}; 
\item
Chemical synthesis process control (reactor settings, safety monitoring); 
\item
Industrial welding control (parameter optimization and quality inspection); 
\item
Path planning (robot motion path generation via ROS\cite{ros} on Gazebo\cite{Gazebo}); 
\item
Network traffic analysis (Wireshark-based inspection and diagnostic reporting)\cite{Wireshark}.
\end{enumerate}

\begin{figure}[t]
\captionsetup{font={footnotesize}}
\begin{center}
\includegraphics[width=\textwidth]{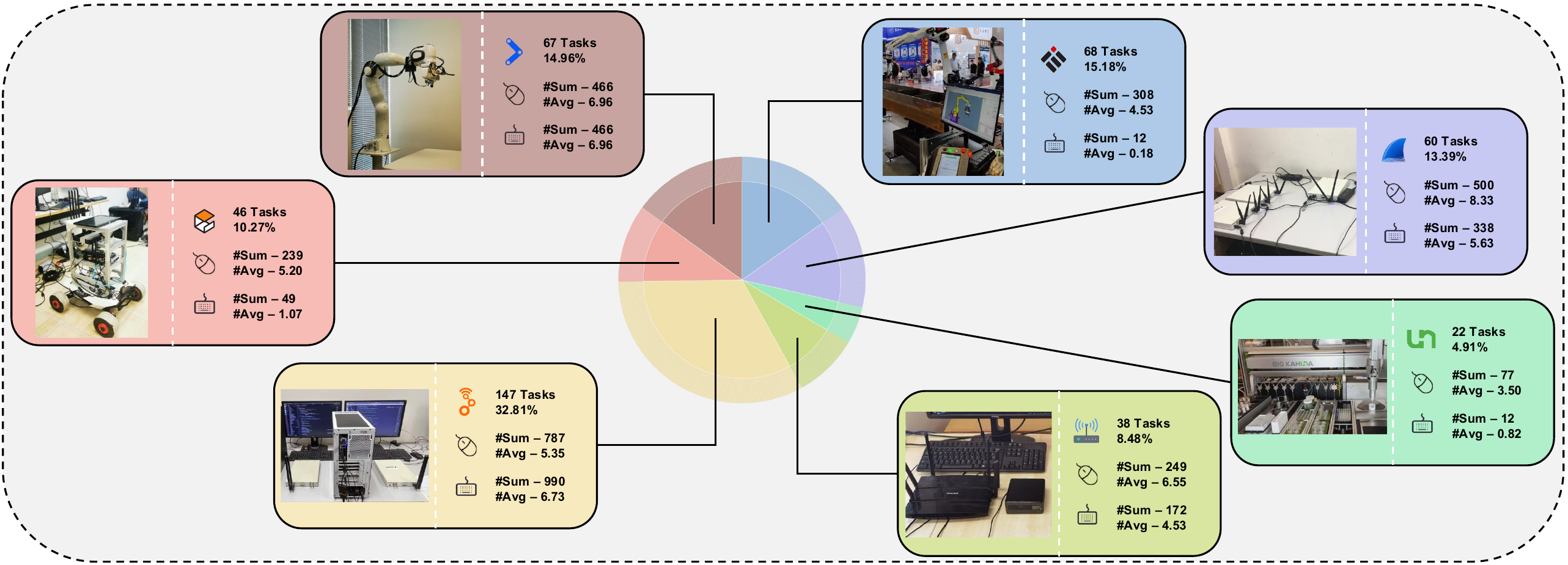}
\caption{Dataset Overview and Statistics. Each subset presents the associated industrial equipment, industrial control software, and statistical information.\vspace{-0.4cm}
}
   \label{fig:dataset}
\end{center}
\vspace{-0.1cm}
\end{figure}

Since IndusGCC is designed to capture the fine-grained visual and behavioral diversity of industrial GUIs, while retaining mission-critical operational details, we recorded the complete process of the human-performed task in a real-world software environment, including mouse movements, clicks, and keyboard inputs. Fig. \ref{fig:dataset} presents the statistics of IndusGCC, including task quantity proportion and average number of steps per task across the seven domains.

\begin{figure}[t]
\captionsetup{font={footnotesize}}
\begin{center}
\includegraphics[width=0.98\textwidth]{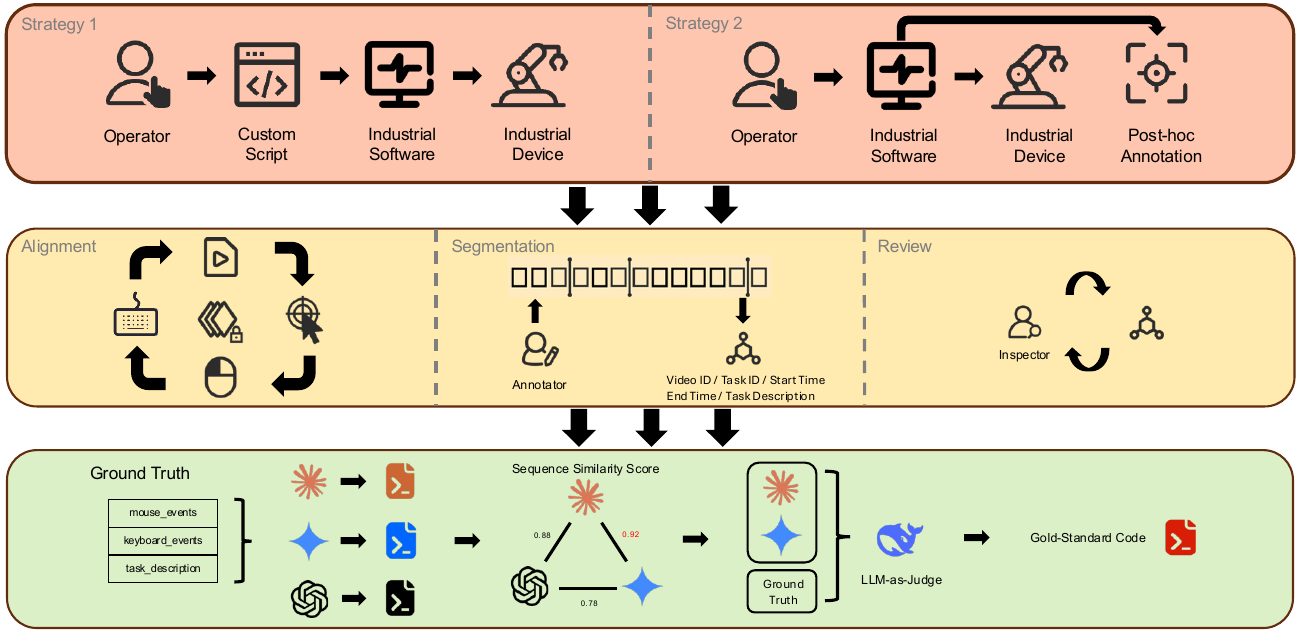}
\caption{An overview of the three-step procedure introduced in Sections \ref{sec:datacollect}, \ref{sec:taskseg}, and \ref{sec:codegen}.\vspace{-0.3cm} }
   \label{fig:dataprocess}
\end{center}
\end{figure}

We also demonstrated how we preprocessed the raw data, including data collection, task division and ground truth code generation (see Fig. \ref{fig:dataprocess} for an illustration of the whole process). Section \ref{sec:datacollect} \ref{sec:taskseg} and \ref{sec:codegen}below introduce the three-step procedure in detail.

\vspace{-0.225cm}\subsection{Data Collection}\vspace{-0.175cm}
\label{sec:datacollect}
During the data collection process, we found that some industrial equipment provided root access to the system. This allowed us to utilize a custom recording script to capture both system logs and screen recordings during workers’ operations. However, other equipment did not grant root access, meaning we were unable to retrieve system logs for analyzing mouse clicks or keyboard inputs. Nevertheless, we were still able to obtain screen recordings of workers’ interactions with the machine, and then we manually annotated the mouse-clicking or keyboard-inputting behaviors of the video to maintain the consistency of data structure. It is worth noting that the mouse and keyboard interaction data were collected solely for generating gold-standard control scripts, which serve as benchmarks for evaluating LLM-GCC performance. These data are not used as inputs when IndusGCC is used for model testing. Due to page limit, implementation details of data collection are given in Appendix \ref{App_B}.

\vspace{-0.225cm}\subsection{Task Segmentation}\vspace{-0.175cm}
\label{sec:taskseg}
Since a single recording session could encompass multiple distinct operational goals, segmenting the raw recordings into well-defined tasks was essential for creating a consistent and reusable dataset. Ideally, each task in our dataset functioned as an independent unit with a clearly defined start point, end point, and operational objective. To achieve this, the construction process of IndusGCC followed a three-phase workflow:

\textbf{Alignment:} All raw recordings underwent a rigorous alignment process. Video frames, mouse events, mouse positions, and keyboard events were synchronized based on frame indices, ensuring a frame-level lock across all modalities. This guaranteed that every task segment contained perfectly matched visual and interaction data, preventing temporal drift between modalities. The alignment process also ensured that when tasks were later segmented, each segment remained internally consistent and ready for step-by-step reproduction.

\textbf{Segmentation:} After recording, each video was assigned to a member of the annotation team who was familiar with the equipment operations depicted in the video. The annotator segmented the video into multiple tasks according to 1) a unified set of guidelines to ensure dataset-wide consistency, and 2) their expertise in the specific equipment’s operation. For each identified task, the annotator created a structured task description that included 1) the video ID, 2) the task ID, 3) the start and end timestamps, and 4) the task description. The task description consisted of two key elements: the overall task goal and a step-by-step, high-level description of the control interactions within the task.

\textbf{Review:} The segmented tasks and their detailed descriptions were reviewed by another team member with expertise in the relevant machine operations. This review ensured that: 1) task boundaries were technically valid, 2) execution steps were complete and accurate, and 3) task semantics aligned with real industrial workflows.

\vspace{-0.225cm} \subsection{Ground Truth Code Generation}\vspace{-0.175cm}
\label{sec:codegen}
To establish a reliable code reference for model evaluation, this paper, positioned as a dataset and benchmark effort, adopts a multi-agent collaboration framework to generate ground truth for each operation video on industrial equipment. The framework consists of two phases: Cooperative Code Generation and LLM as a Judge.

In the Cooperative Code Generation phase, three LLMs -- GPT-4o, Gemini-2.5-Pro, and Claude 4 Sonnet -- were independently tasked with generating executable control script based on the following inputs: mouse positions, mouse events, keyboard events, and task descriptions. Each model produced a full code script intended to replicate the recorded user operation.

After independent code generation at each LLM, we then analyze the similarities between resulting codes to find the most consistent pair of codes. This is motivated by our observation that if two LLMs reach agreement on the operation sequence (i.e., mouse clicking and keyboard input), then the resulting code is very likely to be the correct one.  To identify the most consistent pair of candidate solutions, pairwise code similarity scores were computed for all possible model pairs. The pair with the highest score was selected as the primary reference set for the next phase.

For the computation of pairwise code similarity, we employ the Smith-Waterman local alignment algorithm \cite{smithwaterman} as in previous works \cite{swcode1,swcode2}. The algorithm measures the structural correspondence between operation sequences by identifying optimally matching subsequences rather than enforcing a global alignment. This ensures that partial but functionally critical overlaps in operation paths are preserved in the scoring process \cite{swcode1}.

Specifically, let the operating sequence of one control script be $A_T = (a_1^T, a_2^T,\ldots,a_m^T)$ and the other be $A_S = (a_1^S,a_2^S,\ldots,a_n^S)$, in which each element within the sequence represents a mouse clicking operation or a keyboard inputting operation. The Smith-Waterman algorithm constructs a dynamic programming matrix $H$ where $H_{i,j}$ represents the highest local alignment score between the prefixes $A_T$ and $A_S$. The recurrence relation is given by:
\begin{align} 
H_{i,j} = max \left \{\begin{matrix} 0 , \\ H_{i - 1 , j - 1} + score (a_{i}^{T} , a_{j}^{S}) , \\ H_{i - 1 , j} - \delta , \\ H_{i , j - 1} - \delta \end{matrix} \right. 
\end{align} 
where$score(a_i^T,a_j^S)$ is the substitution score:
\begin{align} 
score(a_i^T,a_j^S)= \left \{ \begin{matrix} 2,a_i^T = a_j^S \\ -1,a_i^T \neq a_j^S \end{matrix} \right.
\end{align} 
and $\delta$ is the gap penalty, which $\delta = -1$.\\
The algorithm initializes $H_{i,0}=H_{0,j}=0$ and the similarity score is taken as:
\begin{align} 
SimilarityScore\left(A_T,A_S\right)=\frac{max_{\left(i,j\right)}H_{i,j}}{n\ast score\left(a_i^T,a_j^S\right)|_{a_i^T\neq a_j^S}}
\end{align} 
In our context, $A_T$ and $A_S$ correspond to sequences of atomic operations extracted from the execution traces. The final Smith–Waterman score is normalized by the length of the reference sequence to ensure comparability across tasks.

In the second phase, LLM as a Judge, a fourth high-performing model, DeepSeek-V3, was introduced. DeepSeek-V3 was provided with all inputs available to the first-stage models, along with the two most similar code sequences from the previous phase and their similarity comparison results. Using this comprehensive context, DeepSeek-V3 synthesized a refined code sequence that combined the strengths of both candidates while resolving any inconsistencies.

This two-phase LLM collaboration framework aims to minimize potential biases associated with any single model. The resulting code sequences were then reviewed by human experts to ensure correctness and operational equivalence within the annotated clickable regions. The validated outputs were ultimately published as the gold-standard reference code for the corresponding task segments.

\vspace{-0.2cm} \section{Evaluation}\vspace{-0.2cm} 
\subsection{Evaluation Metrics} \vspace{-0.2cm} 
\label{sec:evalmetric}
We compare the machine control script generated by the tested model (TestCode) with the gold-standard code produced through the pipeline described in Section \ref{sec:codegen} (SampleCode). Evaluation is performed across four complementary dimensions designed to capture both semantic correctness and structural fidelity: 1) Task Success Rate, 2) Sequence Similarity Score, 3) Operation Hit Rate, and 4) Operation Redundancy Rate.

\textbf{Task Success Rate:} The first dimension measures whether TestCode and SampleCode are functionally equivalent in accomplishing the intended task. To evaluate this, we employ GPT-4o as an automatic judge: it receives both scripts together with the task description and determines whether they can be regarded as semantically equivalent. This metric reflects end-to-end task success, focusing on functional correctness rather than surface-level similarity.

\textbf{Sequence Similarity Score:} The second dimension evaluates how closely the structural execution flow of TestCode matches that of SampleCode. We compute this by aligning the two operation sequences using a local sequence alignment algorithm. Intuitively, the algorithm rewards matching operations, penalizes mismatches and gaps, and ultimately identifies the best-matching subsequences. The final score reflects the degree of overlap in execution paths, normalized by task length to allow fair comparison across tasks.

\textbf{Operation Hit Rate:} The third dimension focuses on parameter-level accuracy for critical GUI operations. We extract mouse-related events (e.g., \textit{click(), moveTo(), dragTo()}) and keyboard-related events (e.g., \textit{typewrite(), press(), keyDown()}). A mouse event in TestCode is counted as correct if its coordinates fall within the annotated tolerance region of the corresponding SampleCode event. For keyboard events, only exact string matches are accepted. The hit rate is then calculated as the proportion of correctly executed operations relative to all required operations, thereby quantifying precision in reproducing human interactions.

\textbf{Operation Redundancy Rate:} Finally, we assess whether the generated code contains unnecessary steps beyond those required for successful execution. Redundancy is defined as the proportion of extraneous operations in TestCode relative to the number of valid steps. This rate can be either positive, indicating that TestCode introduces redundant operations compared to the gold-standard script, or negative, suggesting that TestCode omits essential steps. Values closer to zero imply minimal redundancy, reflecting a more faithful and efficient reproduction of the intended procedure. 

\vspace{-0.2cm} \subsection{Evaluation Framework} \label{sec:evalframe}
\begin{figure}[t]
\begin{center}
\includegraphics[width=\textwidth]{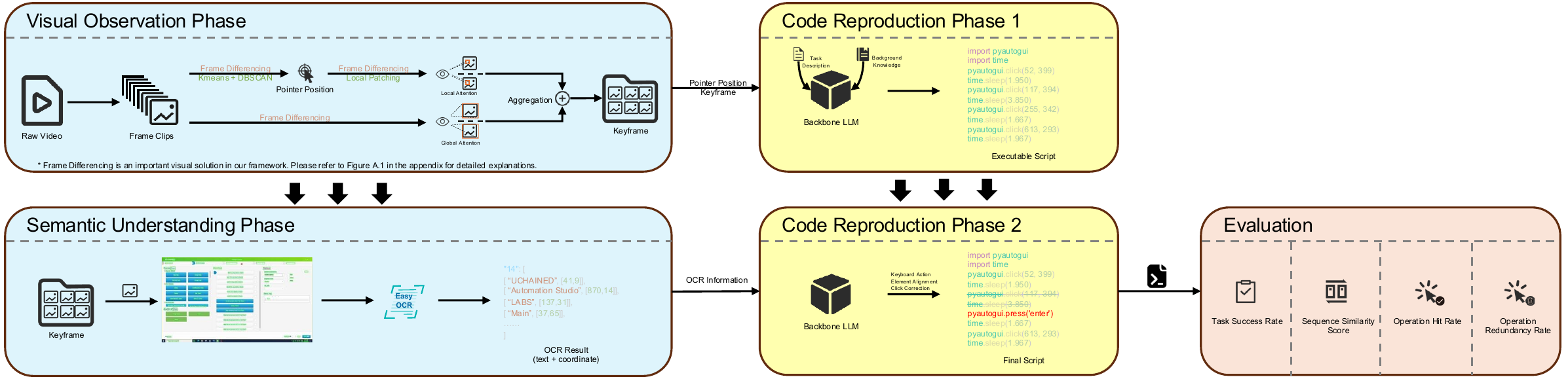}
\vspace{-0.3cm}
\captionsetup{font={footnotesize}}
\caption{An illustration of the evaluation framework. \vspace{-0.3cm}}
\label{fig:eval}
\end{center}
\vspace{-0.3cm}
\end{figure}

\vspace{-0.2cm} 
To realize fully automatic control script generation based on video inputs, specialized tools have been developed to better understand the video input \cite{Chemist-X}. These tools can precisely identify key operation frames from the given raw video and conduct optical character recognition (OCR) for each identified frame to extract operation information, significantly reducing the workload of later LLM analysis. While the ultimate goal of LLM-GCC is to develop powerful AIs that are able to address the video-code transformation with a single model, this paper, as a dataset and benchmarking effort, validates the feasibility of this line of research with the cooperation of multiple models. Specifically, our later experiments leveraged the tool open-sourced in \cite{Chemist-X} as pre-processing (see those highlighted in blue in Fig. \ref{fig:eval}), and we then used text-based models to generate executable Python scripts restricted to the PyAutoGUI API \cite{pyautogui} for GUI control (see those highlighted in yellow in Fig. \ref{fig:eval}). This was followed by an evaluation stage that considered all evaluation metrics described in Section \ref{sec:evalmetric} (see those highlighted in red in Fig. \ref{fig:eval}). Due to space constraints, we do not present further explanation of the evaluation framework. We refer interested readers to Appendix \ref{App_A} for details.

\vspace{-0.3cm} 
\section{Experiment and Discussion} \label{sec:experiments}\vspace{-0.3cm} 
\textbf{Experimental Results:}
The evaluation framework introduced above enables text-based LLMs to interpret GUI interfaces and generate executable operation paths without native visual capabilities. We benchmarked eight mainstream text-based models across diverse industrial GUI scenarios. Experimental results indicate that none of these models can serve as a plug-and-play solution for perfectly operating industrial GUI systems. Even the best-performing model, GPT-4 on the Chem subset, achieves only 22.73\% task coverage, highlighting the considerable gap between current LLM capabilities and the requirements of real-world industrial applications.

On sequence-level structural fidelity, we observed a more uniform trend: claude-sonnet-4 clearly dominated, achieving the highest similarity scores in nearly all subsets. This indicates that claude-sonnet-4 succeeds in both producing functionally correct scripts and closely reproducing the canonical sequence of human operations. The alignment highlights the model’s capacity to capture task-specific procedural structures rather than merely generating semantically equivalent but divergent solutions.

A particularly interesting pattern emerges when comparing fine-grained operation accuracy with redundancy. Across most subsets, higher hit rates correlate with increased redundancy, suggesting that models often insert additional operations as a ``safety margin” to maximize coverage of effective interaction zones, even though introduce inefficiency and potential error propagation. Notably, claude-sonnet-4 on the AP subset and deepseek-v3 on the Chem subset stand out as rare cases where high accuracy is achieved without inflating redundancy. These results point to promising directions for balancing correctness with efficiency in future model design.

\begin{table}[t]
\caption{Experiment results of Task Success Rate, Sequence Similarity Score,Operation Hit Rate and Operation Redundancy Rate on mainstream models}
\small
\setlength{\tabcolsep}{0.6mm}{
\begin{tabular}{p{1.6cm}|l|ccccccc}
\multirow{2}{*}{Metrics}                   & \multirow{2}{*}{Model}   & \multicolumn{7}{c<{\centering}}{Data Subset}                                                                                                        \\ \cline{3-9} 
                                           &                          & AP               & Chem             & MOVEIT           & ROS              & USRP             & Weld             & Wireshark \rule{0pt}{1.0em}     \\ \hline
\multirow{8}{1.6cm}{Task Success Rate}         & claude-sonnet-4-20250514 & \textbf{7.90\%}  & 18.18\%          & \textbf{14.93\%} & 6.52\%           & 4.76\%           & 1.47\%           & 0.00\%          \\
                                           & llama3.1-405b            & 5.26\%           & 9.09\%           & 7.58\%           & 0.00\%           & 18.75\%          & 1.47\%           & 0.00\%          \\
                                           & gpt-4o                   & 2.63\%           & 13.64\%          & 8.96\%           & 2.17\%           & \textbf{19.18\%} & 2.94\%           & 1.67\%          \\
                                           & grok-3                   & 5.26\%           & 18.18\%          & 4.48\%           & \textbf{8.70\%}  & 5.44\%           & \textbf{4.41\%}  & 0.00\%          \\
                                           & deepseek-v3-250324       & 0.00\%           & 13.64\%          & 5.97\%           & 0.00\%           & 7.59\%           & 2.94\%           & 1.67\%          \\
                                           & doubao-1.5-pro-256k      & 0.00\%           & 0.00\%           & 5.97\%           & 0.00\%           & 5.48\%           & 2.94\%           & \textbf{3.33\%} \\
                                           & gpt-4                    & 2.63\%           & \textbf{22.73\%} & 5.97\%           & 2.17\%           & 17.81\%          & 1.47\%           & \textbf{3.33\%} \\
                                           & kimi-k2-instruct         & 2.63\%           & 9.09\%           & 7.46\%           & 4.35\%           & 8.90\%           & 2.94\%           & 1.67\%          \\ \hline
\multirow{8}{1.6cm}{Sequence Similarity Score} & claude-sonnet-4-20250514 & \textbf{61.50\%} & \textbf{73.42\%} & \textbf{63.99\%} & 60.35\%          & \textbf{38.31\%} & \textbf{84.13\%} & 48.78\%          \\
                                           & llama3.1-405b            & 33.63\%          & 54.33\%          & 40.32\%          & 48.07\%          & 29.05\%          & 56.52\%          & 42.96\%          \\
                                           & gpt-4o                   & 35.91\%          & 59.05\%          & 40.92\%          & 50.99\%          & 27.18\%          & 59.29\%          & 35.27\%          \\
                                           & grok-3                   & 35.41\%          & 63.59\%          & 46.33\%          & 48.19\%          & 28.20\%          & 63.83\%          & 39.73\%          \\
                                           & deepseek-v3-250324       & 56.94\%          & 70.35\%          & 60.38\%          & \textbf{69.30\%} & 35.62\%          & 80.80\%          & \textbf{52.88\%} \\
                                           & doubao-1.5-pro-256k      & 40.52\%          & 52.80\%          & 47.42\%          & 44.69\%          & 25.63\%          & 62.16\%          & 43.10\%          \\
                                           & gpt-4                    & 34.94\%          & 56.87\%          & 41.54\%          & 50.12\%          & 26.59\%          & 57.80\%          & 36.30\%          \\
                                           & kimi-k2-instruct         & 56.15\%          & 64.77\%          & 48.23\%          & 58.55\%          & 29.84\%          & 78.11\%          & 48.77\%          \\ \hline
\multirow{8}{1.6cm}{Operation Hit Rate}        & claude-sonnet-4-20250514 & \textbf{65.65\%} & 58.32\%          & 79.10\%          & 66.01\%          & 45.69\%          & 44.95\%          & 39.33\%          \\
                                           & llama3.1-405b            & 61.70\%          & 54.80\%          & 74.24\%          & 44.93\%          & 49.94\%          & 48.18\%          & 44.73\%          \\
                                           & gpt-4o                   & 58.63\%          & 60.59\%          & 83.58\%          & 55.80\%          & \textbf{52.11\%} & \textbf{51.12\%} & 47.13\%          \\
                                           & grok-3                   & 65.01\%          & 58.89\%          & \textbf{86.57\%} & \textbf{73.84\%} & 49.61\%          & 48.38\%          & \textbf{49.71\%} \\
                                           & deepseek-v3-250324       & 59.55\%          & \textbf{63.43\%} & 83.58\%          & 62.71\%          & 46.81\%          & 40.76\%          & 41.13\%          \\
                                           & doubao-1.5-pro-256k      & 58.15\%          & 36.48\%          & 58.21\%          & 38.41\%          & 38.85\%          & 33.97\%          & 34.55\%          \\
                                           & gpt-4                    & 59.07\%          & 61.16\%          & 85.32\%          & 54.27\%          & 51.25\%          & 46.94\%          & 43.70\%          \\
                                           & kimi-k2-instruct         & 61.70\%          & 57.07\%          & 82.34\%          & 67.32\%          & 44.21\%          & 49.36\%          & 41.20\%          \\ \hline
\multirow{8}{1.6cm}{Operation Redundancy Rate} & claude-sonnet-4-20250514 & \textbf{0.63\%}  & 24.85\%          & 8.16\%           & \textbf{7.34\%}  & -26.80\%         & 49.03\%          & 29.36\%          \\
                                           & llama3.1-405b            & 20.54\%          & 28.82\%          & 9.75\%           & 45.22\%          & 28.55\%          & 44.50\%          & 29.85\%          \\
                                           & gpt-4o                   & 14.83\%          & 16.44\%          & -14.59\%         & 29.38\%          & 16.98\%          & 41.64\%          & \textbf{2.18\%}  \\
                                           & grok-3                   & 20.80\%          & 20.56\%          & 21.97\%          & 22.61\%          & 23.79\%          & 50.28\%          & 18.15\%          \\
                                           & deepseek-v3-250324       & 17.69\%          & \textbf{9.81\%}  & 8.07\%           & 16.00\%          & 24.65\%          & 46.69\%          & 38.41\%          \\
                                           & doubao-1.5-pro-256k      & 26.53\%          & 26.25\%          & 18.93\%          & 31.48\%          & 19.37\%          & \textbf{33.69\%} & 11.11\%          \\
                                           & gpt-4                    & 18.46\%          & 13.26\%          & -7.06\%          & 36.81\%          & \textbf{14.42\%} & 50.54\%          & 20.09\%          \\
                                           & kimi-k2-instruct         & 15.88\%          & 17.79\%          & \textbf{6.87\%}  & 7.53\%           & 19.33\%          & 36.20\%          & 37.06\%         
\end{tabular}
}
\vspace{-0.6cm}
\normalsize
\end{table}

\textbf{Observations and Discussions:} From an algorithmic perspective, the suboptimal performance of the current framework mainly stems from its limited ability to interpret fine-grained visual information in GUI environments. The reliance on a zero-shot frame-differencing method restricts context adaptation, leading to imprecise recognition of pointer positions, icon shapes, subtle color changes, or menu states. A promising solution is to fine-tune a tool-oriented vision encoder trained on GUI screenshots or adopt multimodal large models that jointly process textual prompts and visual signals, improving alignment between perceived states and operational actions. Moving from zero-shot to few-shot or domain-adapted fine-tuning may also help internalize common interaction patterns and enhance task success rates.

From a dataset perspective, the current benchmark covers a constrained range of GUI operations, limiting generalization to diverse scenarios. Expanding it with richer manipulations -- such as viewport rotation, zooming, scrolling, drag-and-drop, and shortcut key combinations -- would create a more representative evaluation space and foster the development of models capable of handling complex, realistic human–computer interaction tasks.

\vspace{-0.2cm}\section{Conclusion}\vspace{-0.2cm}
This work presents the first general computer control (GCC) dataset for industrial scenarios, covering diverse applications, such as robotics control, mission-critical industrial networking, and automatic chemical synthesis. We design automated data collection pipelines coupled with professional verification to ensure both scalability and quality. Building upon this dataset, we introduce a comprehensive evaluation framework that defines four complementary metrics, i.e., LLM-judged success rate, AST similarity, operation hit rate, and redundancy rate, to enable fine-grained, end-to-end assessment of model performance on industrial GCC tasks. Our benchmarking results across mainstream models reveal their promising capabilities and current limitations, highlighting persistent challenges in accurate operation sequencing and efficiency. This work serves as a preliminary effort in building the domain-specific dataset and calls for further investigation building upon the proposed benchmarks.

\vspace{-0.2cm} \section{Acknowledge}\vspace{-0.2cm}
We acknowledge our industrial partners for providing the real-world factory data used in this work.

\clearpage

\bibliographystyle{unsrt}
\bibliography{reference}

\newpage
\appendix
\section{Algorithm details of Evaluation Framework}

\textbf{Visual Observation Phase}
In GUI-based industrial software operations, the mouse pointer or the input cursor serves as the primary locus of human visual attention. We define key events as either mouse clicks or keyboard inputs, and the corresponding video frames in which these events occur as key frames.

We found that most key events on industrial GUIs exhibit “trigger-and-feedback” patterns—either local changes (e.g., button highlighting, text field activation) or global changes (e.g., page refresh or redrawing) upon interaction. By measuring both local and global pixel-level differences between consecutive frames, the framework can detect significant state transitions and pinpoint the mouse pointer location at event time. This hybrid approach is inspired by human selective attention, and we term it the Global–Local Attention.

\begin{figure}[htbp]
\begin{center}
   \includegraphics[width=\textwidth]{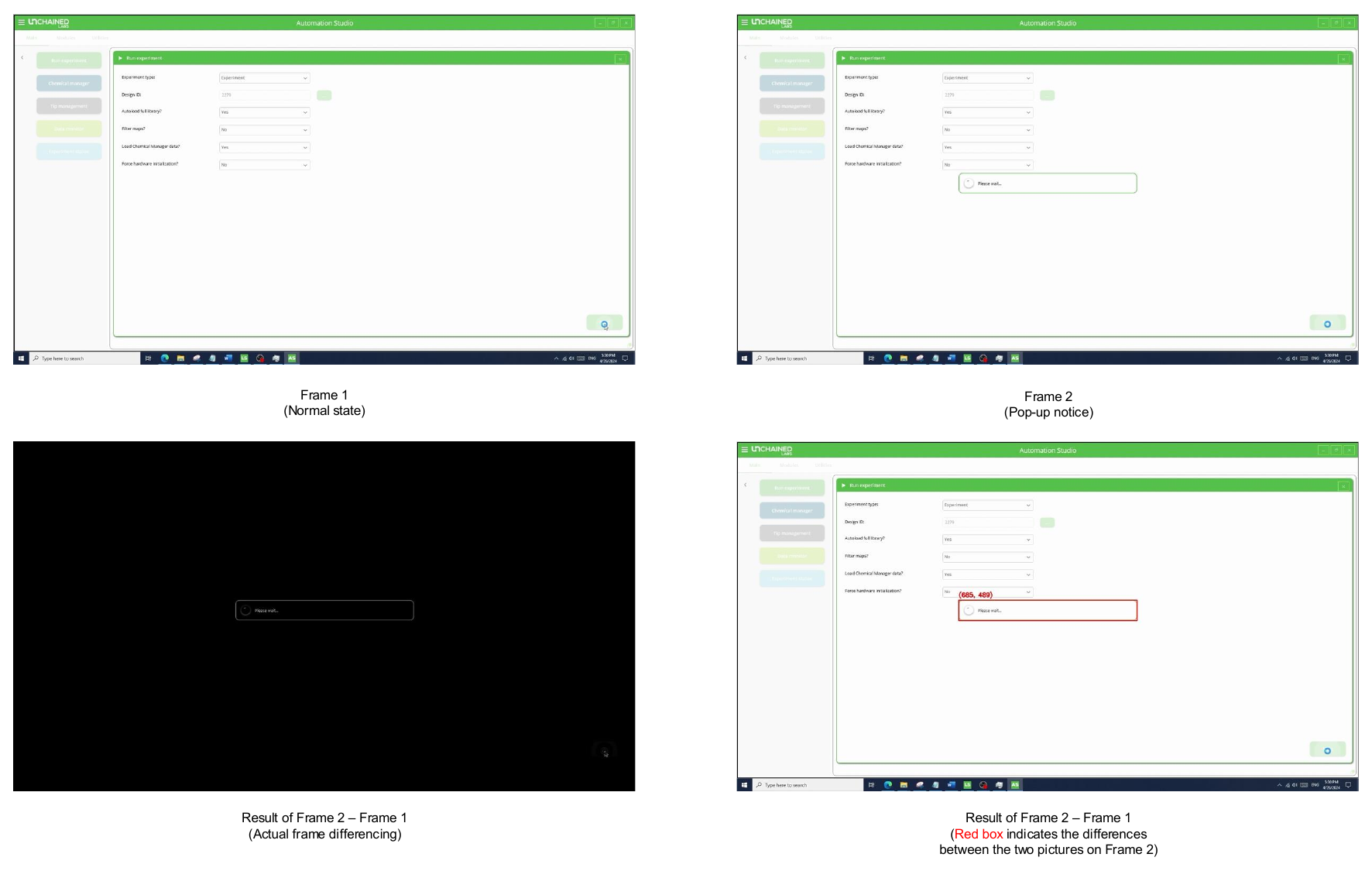}
\vspace{-0.3cm}
\caption{An example demonstrating Frame Differencing method}
   \label{fig:framediff}
\end{center}
\vspace{-0.1cm}
\end{figure}

To automatically endow the model with the Global–Local Attention capability to achieve automatic detection of mouse pointers and keyframes, we employ a Frame Differencing method, which computes the absolute difference of each pixel between two consecutive frames to quantify the degree of visual change. As illustrated in Figure \ref{fig:framediff}, Frame 2 introduces a new popup window compared with Frame 1. By rendering the result, the non-overlapping regions become visible, clearly highlighting the new elements (in this case, located inside the red box on Frame 2). 

\textbf{Semantic Understanding Phase}
While modern LLMs possess strong language understanding capabilities, their performance in spatial grounding remains limited. To bridge this gap, we integrate Optical Character Recognition (OCR) into the workflow. Using the EasyOCR library, we extract textual content from each key frame, along with the corresponding bounding-box coordinates. This enables a joint semantic–spatial representation of the GUI state, facilitating accurate mapping between textual UI elements and their screen positions in downstream reasoning steps.

\textbf{Code Reproduction Phase}
Code generation proceeds in a two-stage chat with the LLM backbone:

Initial Code Drafting — The model is provided with the dataset background, task goal description, and a sequence of detected key frames along with predicted mouse click coordinates. The model is instructed to output a first-pass executable script using only pyautogui functions for GUI manipulation.

Code Refinement — The model receives the OCR-derived semantic–spatial annotations of key frames and is tasked with refining the initial script. This may include inserting missing keyboard actions, correcting mouse click coordinates and aligning interactions with text-labeled interface elements.

The final output of this two-stage generation process is an executable Python program, which is subsequently compared against the gold-standard SampleCode using the evaluation metrics from Section \ref{sec:evalmetric}.
 \label{App_A}
\section{Implementation details of Data Collection}
We now detail our data collection process for the two scenarios mentioned in Section \ref{sec:datacollect} as follows.

\textbf{Strategy One:} Script-Based Recording. For equipment with root access, we used the script open-sourced alongside the dataset at https://github.com/Golden-Arc/IndustrialLLM. Once initiated via a terminal, the script continuously captured:

1) \textit{Full-Screen Video}: Recorded at a resolution of 1920×1080 and a frame rate of 60 FPS, utilizing FFmpeg for efficient real-time encoding.

2) \textit{Mouse Movement Trajectory}: Captured as tuples of (timestamp, x-coordinate, y-coordinate) for each frame, ensuring sub-frame temporal resolution.

3) \textit{Mouse Events}: Logged with timestamp, x-coordinate, y-coordinate, button type (e.g., left/right), and button state (e.g., pressed/released).

4) \textit{Keyboard Events}: Recorded with timestamp, key identifier, and key states (e.g., pressed/released).

The script is designed for stable operation on commonly used operating systems where equipment control software is deployed (such as the latest Ubuntu Release and Windows 10/11) to ensure cross-platform reliability for the LLM-GCC agent. Operators must manually start the script from a command-line terminal and terminate it by entering an interrupt command.

To ensure consistent and high-quality data collection, each equipment operator underwent a structured training session and followed a step-by-step tutorial prior to recording. Operators conducted three preliminary trial runs to confirm environment compatibility, validate accurate script output, and ensure adherence to dataset specifications. During the formal recording phase, operators were not provided with pre-defined task scripts; instead, they were encouraged to perform operations that reflect real-world industrial production tasks, particularly those involving machine configuration. To maintain data clarity and reproducibility, operators were instructed to minimize errors and redundant actions, ensuring that recorded sequences are concise, executable, and representative of practical use cases.

\textbf{Strategy Two:} Post-Hoc Manual Annotation. In certain industrial environments, running the recording script was not feasible due to restricted system access. In such cases, we resorted to manual post-hoc annotation based on the screen recording videos. Specifically, each video was first processed at a standard frame rate of 10 FPS and a resolution of 1920×1080. The videos were then decomposed into continuous frames using OpenCV, and each frame was annotated with the VGG Image Annotator (VIA), a lightweight, open-source annotation tool developed by the Visual Geometry Group \cite{dutta2016via,dutta2019vgg}. To ensure consistency across data acquisition modes, the annotation schema for manual labeling mirrored that of the automated recording script. Annotators were instructed to label the following information for each frame: 1) mouse position coordinates, including timestamp and the location of the mouse; 2) mouse click events, including button type and mouse state; 3) keyboard events, including timestamp, key identifier, and the state of the keyboard. \label{App_B}

\end{document}